\documentclass[manuscript,authorversion,nonacm]{acmart}

\DeclareMathDelimiter{(}{\mathopen} {operators}{"28}{largesymbols}{"00}
\DeclareMathDelimiter{)}{\mathclose}{operators}{"29}{largesymbols}{"01}

\usepackage[latin1]{inputenc}
\usepackage[T1]{fontenc}

\usepackage{graphicx}
\usepackage{enumerate}
\usepackage{float}
\everymath{\displaystyle}

 \usepackage{booktabs}
\usepackage{longtable}
\usepackage{changepage}
\usepackage{lscape} 
\usepackage{multirow}
\usepackage{tabularx}
\usepackage{booktabs}

\usepackage{booktabs}
\usepackage{multirow}
\usepackage[ruled]{algorithm2e}
\SetAlFnt{\algofont}
\SetAlCapFnt{\algofont}
\SetAlCapNameFnt{\algofont}
\SetAlCapHSkip{0pt}
\IncMargin{-\parindent}

\usepackage{hyperref}
\usepackage{graphicx}
\usepackage[skins]{tcolorbox}
\copyrightyear{2019}
\acmYear{2019}
\setcopyright{acmlicensed}
\acmConference[CSCW '19]{CSCW '19: The 22nd ACM Conference on Computer-Supported Cooperative Work and Social Computing}{November 09--13, 2019}{Austin, Texas}
\acmBooktitle{Woodstock '19: The 22nd ACM Conference on Computer-Supported Cooperative Work and Social Computing, November 09--13, 2019, 2019, Austin, Texas}
\acmPrice{15.00}

\begin{document}
\title{Stance Detection on Social Media: State of the Art and Trends}

%

\author{Abeer ALDayel}
\authornotemark[]
\email{a.aldayel@ed.ac.uk}
\affiliation{
  \institution{School of Informatics. The University of Edinburgh, Edinburgh, UK}
}
\author{Walid Magdy}
\authornotemark[]
\email{wmagdy@inf.ed.ac.uk}
\affiliation{
  \institution{School of Informatics. The University of Edinburgh, Edinburgh, UK}
}

%
\renewcommand{\shortauthors}{Abeer Aldayel \& Walid Magdy}
\begin{abstract}
Stance detection on social media is an emerging opinion mining paradigm for various social and political applications in which sentiment analysis may be sub-optimal.
There has been a growing research interest for developing effective methods for stance detection methods varying among multiple communities including natural language processing, web science, and social computing. This paper surveys the work on stance detection within those communities and situates its usage within current opinion mining techniques in social media. It presents an exhaustive review of stance detection techniques on social media, including the task definition, different types of targets in stance detection, features set used, and various machine learning approaches applied. The survey reports state-of-the-art results on the existing benchmark datasets on stance detection, and discusses the most effective approaches. In addition, this study explores the emerging trends and different applications of stance detection on social media. The study concludes by discussing the gaps in the current existing research and highlights the possible future directions for stance detection on social media. 

\begin{tcolorbox}[skin=widget,
boxrule=1mm,
coltitle=red,
colframe=red!25!white,
colback=blue!5!white,
width=(\linewidth),before=\hfill,after=\hfill,
adjusted title={\large{\textbf{Note}}}]
\large{This is an early version of the paper. For the final version, please find it published at Elsevier IP\&M in the following link: \url{https://doi.org/10.1016/j.ipm.2021.102597} . } 
\end{tcolorbox}
\end{abstract}

\maketitle

\section{Introduction}

Nowadays, social media platforms constitute a major component of an individual's social interaction. These platforms are considered robust information dissemination tools to express opinions and share views. People rely on these tools as the main source of news to connect with the world and get instant updates \cite{newman2011mainstream}. They are very beneficial because they allow individuals to explore various aspects of emerging topics, express their own points of view, get instant feedback, and explore the views held by the public. The huge dependency of users on these platforms as their main source of communication allows researchers to study different aspects of online human behavior, including the public stance toward various social and political aspects.

Stance is defined as the expression of the speaker's standpoint and judgement toward a given proposition  \cite{biber1988adverbial}.
Stance detection plays a major role in analytical studies measuring public opinion on social media, particularly on political and social issues. The nature of these issues is usually controversial, wherein people express opposing opinions toward differentiable points.
Social issues such as abortion, climate change, and feminism have been heavily used as target topics for stance detection on social media \cite{mohammad2016}. Similarly, political topics, such as referendums and elections, have always been hot topics that have been used in stance detection to study public opinion \cite{fraisier2018stance}. 
Generally, the stance detection process is also known as perspective \cite{klebanov2010vocabulary,elfardy2017perspective} and viewpoint \cite{zhu2019hierarchical,trabelsi2018unsupervised} detection, in which perspectives are identified by expressing stances toward an object of a controversial topic \cite{elfardy2017perspective}. 

The stance has been used in various research as a means to link linguistic forms and social identities that have the capability to better understand the background of people with polarized stances \cite{bassiouney2015stance}. Consequently, the early work on stance detection emerged in analyzing political debates on online forums 
\cite{lin2006side,somasundaran_recognizing_2009,murakami_support_2010,walker_corpus_2012}.

\begin{figure}
\centering
\begin{minipage}[b]{0.70\textwidth}
    \includegraphics[width=\textwidth]{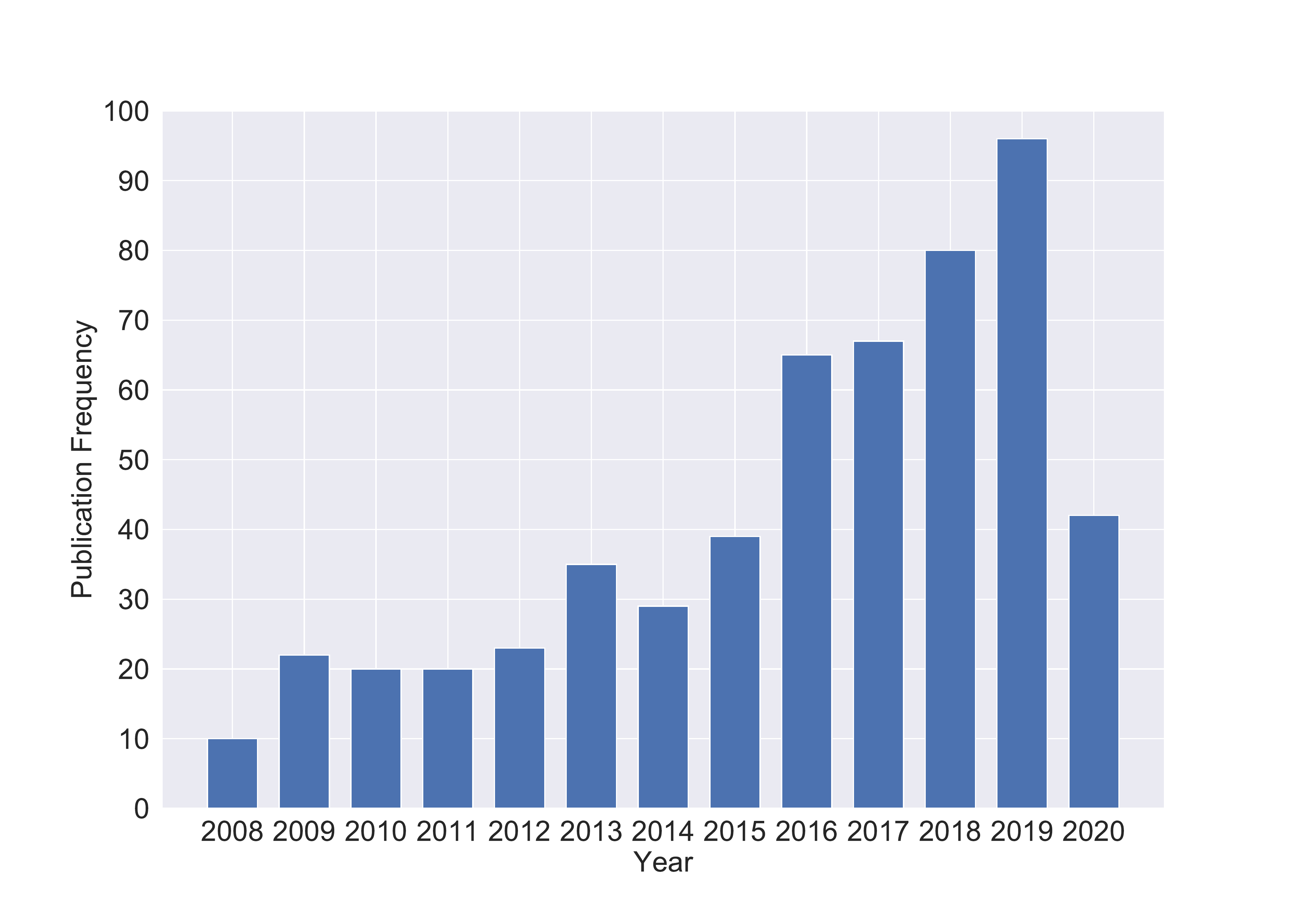}
    \caption{\textcolor{black}{Publications per year on stance detection as searched on Web of Science. The following keywords were used for search: ``stance detection'', ``stance prediction'' and ``stance classification''.}}
    \label{fig:publications}
  \end{minipage}
\end{figure}

Recently, a growing body of work has used the concept of stance detection on social media as a new frontier to analyze various social and political issues across social media platforms. 
Figure~\ref{fig:publications} shows the number of publications from 2008 to 2020 on stance detection. As shown, there is a noticeable steady increase in the number of publications, with a sudden increase from 2016, which we relate to the release of the SemEval 2016 stance detection task \cite{mohammad2016}, which provided the first benchmark dataset for stance detection on social media.
Earlier work on stance detection focused on analysing of debates on online forums, which is distinguished from the more recent work that focused more on social media platforms, especially Twitter, since the former has a clear single context in comparison to social media platforms. In the online forums, the users debate in the form of a thread discussion where there is a flow of information related to the topic. In contrast, social media lacks such representations where the users participate solely with a topic  and rarely with a context (i.e., thread). 

As can be noticed from Figure~\ref{fig:publications}, one of earliest initiatives to promote stance detection on social media is the "SemEval 2016 task 6" shared-task, which introduced a stance-specific
benchmark dataset to help in evaluating stance detection on Twitter \cite{mohammad2016}. With a major focus on identifying stances toward social issues, SemEval-2016 has formulated the stance detection task around a few controversial issues, such as atheism, feminism, and opinions regarding some politicians. In addition, a new wave of stance detection applications has been triggered to handle issues that have infected social media lately, such as fake news and rumors  \cite{gorrell-etal-2019-semeval,derczynski_SemEval-2017_2017,aker_stance_2017}, where the stance toward claims in a piece of news is utilized as a key feature for the detection of the credibility of that news.

\begin{figure}
\centering

\end{figure}


  
The majority of work on stance detection has targeted the detection of the stance toward a given subject expressed in a given text. However, some works have studied the detection of the stances of users toward subjects without explicitly stating them, which is usually referred to as stance prediction.
Thus, the work on stance detection can be categorized into two main types, namely, detecting expressed views versus predicting unexpressed views. In the first type, the objective is to classify a user's post and infer the current stance to be in favor of or against a given subject~\cite{mohammad2016}.

In the latter, the prediction is carried out to infer a user's viewpoint on a given topic that the user has not discussed explicitly or toward an event that has not occurred yet. This type of stance detection has proven its effectiveness in predicting the future attitudes in the aftermath of an event \cite{darwish_predicting_2017,magdy__2016}.

The work on stance detection can also be categorized based on the topic of the target of analysis, that is, it can be one specific target, multiple related targets, or a claim in a news article. Most existing designs of stance detection classifiers work to identify the user's stance toward a specific topic. Sometimes, the classifier is built to detect the stance toward multiple related targets. This is a situation when the stance is detected toward two related entities that are typically opposed to against each other, such as the detection of the stance toward Clinton and Trump simultaneously, since if the stance is in favor one target, it will simply be against the other \cite{sobhani_dataset_2017}. When the target is a claim in a news statement, the main objective is to identify if a given claim is supported by other posts. In this type of detection, the analyses are between two posts, such as in RumourEval~\cite{gorrell-etal-2019-semeval,derczynski_SemEval-2017_2017} or between news headers and other bodies of articles; these are mainly used as an initial step for fake news detection \cite{allcott_social_2017}

 \textcolor{black}{To  our best knowledge, this is the first work to cover the current research directions on stance detection on social media and identify a framework for future research. Recently, there have been a few surveys related to stance detection.  The first survey study  \cite{wang2019survey} emphasized more on surveying the opinion mining methods in general, with a focus on stances toward products. Another study by \cite{survey3} provides a comparative analysis of stance detection models with a sole focus on the textual features only .A more recent survey study \cite{kuccuk2020stance} focused more on surveying the research work that modeled stance detection as a text entailment task using natural language processing (NLP) methods. While, the previous studies surveys the stance concerning one domain, in contract our main objective of this article is to survey the stance detection on social media platforms related to multiple research domains, including NLP, computational social science, and Web science.  More-specifically, this study provides the following contributions:}
 \begin{itemize}
\item \textcolor{black}{This study maps out the terrain of existing research on stance detection and synthesizes how this relates to existing theoretical orientations. It provides a through comparison of stance and sentiment for soci-political opinion mining and show the orthogonal relation of sentiment polarity and stance.}  
\item \textcolor{black}{This survey study provides a broader overview of stance detection methods, covering work that has been published in multiple research domains, including NLP, computational social science, and Web science. }
\item \textcolor{black}{This study surveys the modeling of stances using text, networks, and contextual features that have been overlooked by previous surveys. }
\item \textcolor{black}{Since this survey navigates the stance modeling across multiple domains, this study provides a comprehensive  study investigates the stance detection applications on social media and provide a comprehensive discussion on the future trends and future directions of this field of study. }

\end{itemize}

The rest of the paper is organized into several sections. Section 2 provides a theoretical definition of stance. Section 3 contrasts the differences between stance and sentiment. Sections 4 and 5 summarize the literature on stance prediction and categorize this work according to the types of detected stances (i.e., expressed versus unexpressed) and the target of the stance in the text. Sections 6 and 7 explore the different stance modeling and machine learning approaches for stance classification. Section 8 lists several applications of stance detection, such as social media analyses and fake-news detection. Stance detection resources are listed in Section 9. Finally, the current research trends on stance classification are discussed in Section 10, highlighting the gaps and suggesting the potential future work required in this area. 

\section{Definitions of Stance and Stance Detection}
 Biber and Finegan 1988, define a stance as the expression of the speaker's attitude, standpoint, and judgement toward a proposition  \cite{biber1988adverbial}. 
 
\citeN{du2007stance} argues that stance-taking (i.e., a person taking a polarized stance toward a given subject) is a subjective and inter-subjective phenomenon in which the stance-taking process is affected by personal opinions and non-personal factors such as cultural norms. Stance taking is a complex process related to different personal, cultural, and social aspects. For instance, the political stance-taking process depends on experiential behavior, as stated by ~\cite{mckendrick2014taking}.

The process of detecting the stance of a given person on social media is still in its infancy, as it is still unclear what roles of language and social interaction play in inferring the user's stance. 
  
Stance detection has a strong history in sociolinguistics, wherein the main concern is to study the writer's viewpoint through their text.
Stance detection aims to infer the embedded viewpoint from the writer's text by linking the stance to three factors, namely, linguistic acts, social interactions, and individual identity. Using the linguistic features in stance detection is usually associated with attributes such as  adjectives, adverbs, and lexical items \cite{jaffe2009stance}.

It has been argued that stance taking usually depends on experiential behavior, which is based on previous knowledge about the object of evaluation \cite{mckendrick2014taking}. This strengthens stance detection as a major component in various analytical studies.

\begin{figure}
\centering
  \includegraphics[width=0.32\linewidth]{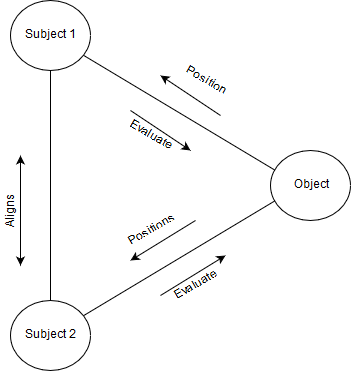}
  \caption{The stance triangle, adapted from  [Du Bois 20]}
  \label{fig:boat1}
\end{figure}

Stance detection in social media is concerned with individuals' views toward objects of evaluation through various aspects related to the users' post and personality traits.

As stated in Due Bois's stance triangle shown in Fig~\ref{fig:boat1}, the process of taking a stance is based on three factors. These are 1) evaluating objects, 2) positioning subject (the self), and 3) aligning with other subjects (i.e., other social actors). For instance, ``\textsf{I am with the new legalization on climate Change}'' has a subject, ''self,'' indicated by proposition  ``\textsf{I}'' and further, the ``\textsf{with}'' indicates the favored position toward the object ```\textsf{Climate Change}''. 
In social media, identifying the stance subject (self) is mostly straightforward as each post is linked to the user.

Furthermore, from a sociolinguistic perspective \cite{jaffe2009stance}, it has been argued that there is no completely neutral stance, as people tend to position themselves through their texts to be in favor of or against the object of evaluation. This creates a further complexity in identifying the stance of the social actors, since a stance is not usually transparent in the text, but sometimes must be inferred implicitly from a combination of interaction and historical context.

\section{Stance vs. Sentiment}
In social media analysis studies, there is a noticeable misconception between sentiment and stance. Some of the studies use sentiment analyzers as the main method to measure the support of a given target, which is sub-optimal \cite{aldayel_assessing_2019}. In this section, we highlight the main theoretical differences between stance and sentiment, and illustrate these differences with some examples. In addition, we apply a quantitative study on data labelled for both sentiment and stance to further demonstrate these differences. 

\subsection{Definition of the Tasks}
\label{sec:definition}
Sentiment analysis is a well-known task in NLP to determine the polarity of emotion in a piece of text.
Generally, this task could be defined as estimating the users' emotional polarity as positive, negative, or neutral \cite{pang2008opinion,jurafsky_speech_2014}. This can be seen in related work on SemEval sentiment tasks SemEval-/(2015, 2016, 2017 and 2020), where the aim of the sentiment analysis task is to determine the polarity toward an aspect of a product, such as a cell phone \cite{patwa2020sentimix,pontiki2015SemEval,nakov-etal-2016-semeval,pontiki2016semeval,rosenthal-etal-2017-SemEval}. 
Unlike sentiment analysis, stance detection mainly focuses on identifying a person's standpoint or view toward an object of evaluation, either to be in favor of (supporting) or against (opposing) the topic . This is usually inferred by a mixture of signals besides the linguistic cues, such as the user's feelings, personality traits, and cultural aspects \cite{biber1988adverbial}. 
Sentiment analysis is mostly approached as a linguistic agnostic task that focuses on leveraging a given text's linguistic properties to identify the polarity of the text \cite{benamara_evaluative_2017}, while stance detection can take an additional paradigm by leveraging non-textual features such as networks and contextual features to infer the user's stance  \cite{lahoti2018joint,darwish_improved_2017,aldayel_your_2019}. 

In general, sentiment analysis is concerned with detecting the polarity of the text, which can be inferred without the need to have a given target of interest; for example \textit{"I am happy"}. Thus, the sentiment analysis model can be represented as shown in equation \ref{eq:sentiment_t}, where $T$ is a piece of text, and the outcome is usually one of three labels \{\textit{positive}, \textit{negative}, \textit{neutral}\}. However,  the main sentiment outcome can take different forms, such as binary polarity, multi-level polarity, and regression. 

\begin{equation} \label{eq:sentiment_t}
Sentiment(T) = \{Positive, Negative, Neutral\}
\end{equation}

Another kind of sentiment analysis concerns inferring the polarity of a text by using a predefined set of targets. This kind of sentiment analysis is usually referred to as target-based sentiment analysis \cite{ma2018targeted,pontiki2016semeval,karamibekr2012sentiment,singhsentiment}. In this kind of sentiment analysis, the classification task can be defined as follows:
\begin{equation} \label{eq:sentiment_tg}
\begin{split}
Sentiment(T|G) &= \{Positive, Negative, Neutral\}
\end{split}
\end{equation}

In definition \ref{eq:sentiment_tg}, the input $G$ represents the target or the entity of evaluation.  Still, in this kind of sentiment analysis, the dominant factor to gauge the polarity is the raw text. 

For stance detection, a clear target $G$ must be defined in advance, to assess the overall attitude toward this target.

The basic form of stance detection on social media can be formulated by using the attributes of the social actor. Thus, in this form of stance detection, the main two inputs are 1) text $T$ or user $U$, and 2) given target $G$, as illustrated in equation~\ref{eq:Stance_e}.
\begin{equation} \label{eq:Stance_e}
\begin{split}
Stance(T,U|G) = \{Favor, Against, None\}
\end{split}
\end{equation}

There are three possible variations of this definition, which may include either the text, the social actor,  or both as input for stance detection. The dependent factor in the stance detection task is the \textit{target} of analysis. One of the common practices of stance detection in social media is to infer the stance based on the raw text only, which maps the stance detection problem to a form of textual entailment task \cite{lin2006side,mohammad2016}. According to a more relaxed definition, a text $T$ entails a stance to a target $G$, ($T$ $\longrightarrow$ stance to $G$), if the stance to a target can be inferred from the given text. 

In social media analysis, this formulation has been further extended to include the social actor, which can be represented by a set of online behavioral features to infer the stance. Instead of the dependency on the raw text, the other representation of stance detection includes the social actor $U$ as a salient factor to infer the stance. Reflecting on the origin of stance taking using the stance triangle as defined by Due Bois, as shown in Fig~\ref{fig:boat1} \cite{du2007stance}, the social actor (subject) is the main element of this process. The structure of social media extends the possibility to represent the social actor by using a variety of network features such as profile information (including age and description) \cite{magdy__2016,darwish_improved_2017}.  

To define the interplay between sentiment and stance, several studies have demonstrated that it is insufficient to use sentiment as the only dependent factor to interpret a user's stance  \cite{somasundaran_recognizing_2009,elfardy2016cu,sobhani2016detecting,mohammad_stance_2017,aldayel_assessing_2019}. This is due to the complexity of interpreting stances from a given text, as they are not always directly aligned with the polarity of a given post.  Table \ref{table:Sentiment_Example} shows examples of tweets that illustrate the orthogonal relationship between sentiment polarity and stance.  For instance, in example 1, the tweet is against the legalization of abortion with a positive sentiment.  In this example, the legalization of abortion is the target for which the opinion is expressed. While the text has a positive sentiment, it shows a negative or opposing stance.
Furthermore, a text may not indicate any sentiment and still pose a clear stance toward a target. This is indicated in example 2, where the tweet has a neutral sentiment with an "against" stance toward the legalization of abortion. On the other hand, example 3 shows the opposite case, where a negative sentiment can be detected without a supporting stance toward the target. Finally, in example 4, applying target-dependent sentiment analysis toward Hillary Clinton leads to a clearly negative sentiment, while measuring the stance toward the same target is clearly supportive. These examples demonstrate that both phenomena (sentiment and stance) are orthogonal.

In the following section, a quantitative analysis on the alignment between stance and sentiment toward a set of targets is applied to quantify the sentiment polarity of the expressed stance.

\begin{table}[t]

\caption{sample of tweets illustrating the sentiment polarity of the expressed stance}{%
\begin{tabular} { >{\centering\arraybackslash}m{0.01\linewidth}  >{\centering\arraybackslash}m{0.55\linewidth} >{\centering\arraybackslash}m{0.15\linewidth} >{\centering\arraybackslash}m{0.08\linewidth} >{\centering\arraybackslash}m{0.08\linewidth} }

  \hline

   \# & Tweet & Target & Sentiment & Stance  \\
  \hline\hline

1 & It is so much fun having younger friends who are expecting babies. \#beentheredonethat \#chooselife .  & Legalisation of Abortion & + & -  \\ \hline
2 & Life is sacred on all levels. Abortion does not compute with my philosophy. (Red on \#OITNB ) \.   & Legalization of Abortion & 0 & -  \\ \hline
3 & The  biggest  terror  threat  in  the  World  is  climate  change \#drought \#floods & Climate Change is the real concern &- & +\\ \hline
4 & I am sad that Hillary lost this presidential race & Hillary Clinton& - & + \\

\hline

\end{tabular}
}
\label{table:Sentiment_Example}
\centering
\end{table}

\subsection{Stance vs Sentiment Quantitatively}
\label{sec:senti_analysis}

\begin{figure}
\centering
\begin{minipage}[b]{0.32\textwidth}
    \includegraphics[width=\textwidth]{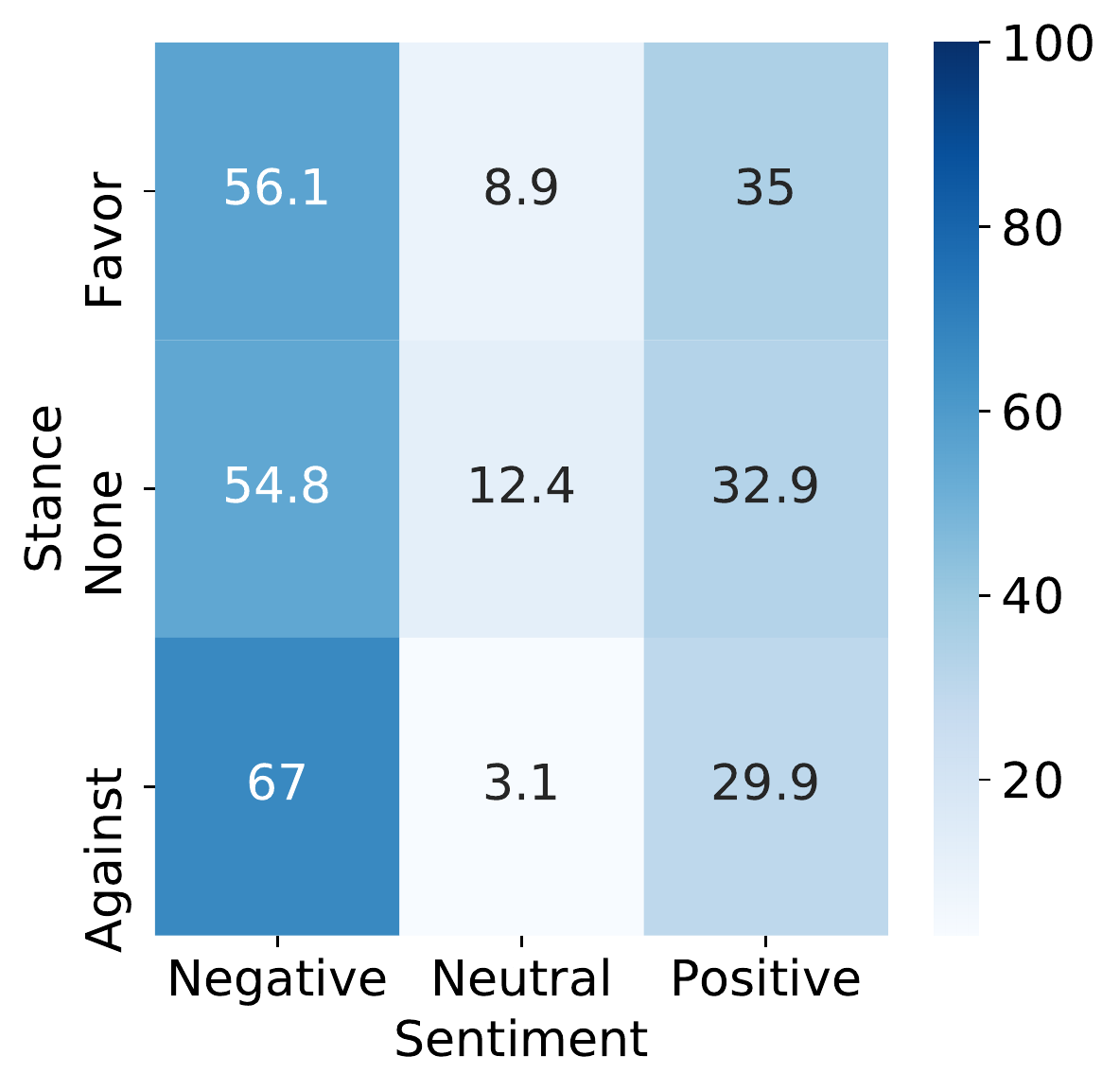}
    \caption{Sentiment distribution over stance.}
    \label{fig:heat_map}
  \end{minipage}
\end{figure}

\begin{figure}
\centering
\begin{minipage}[b]{0.709\textwidth}
    \includegraphics[width=\textwidth]{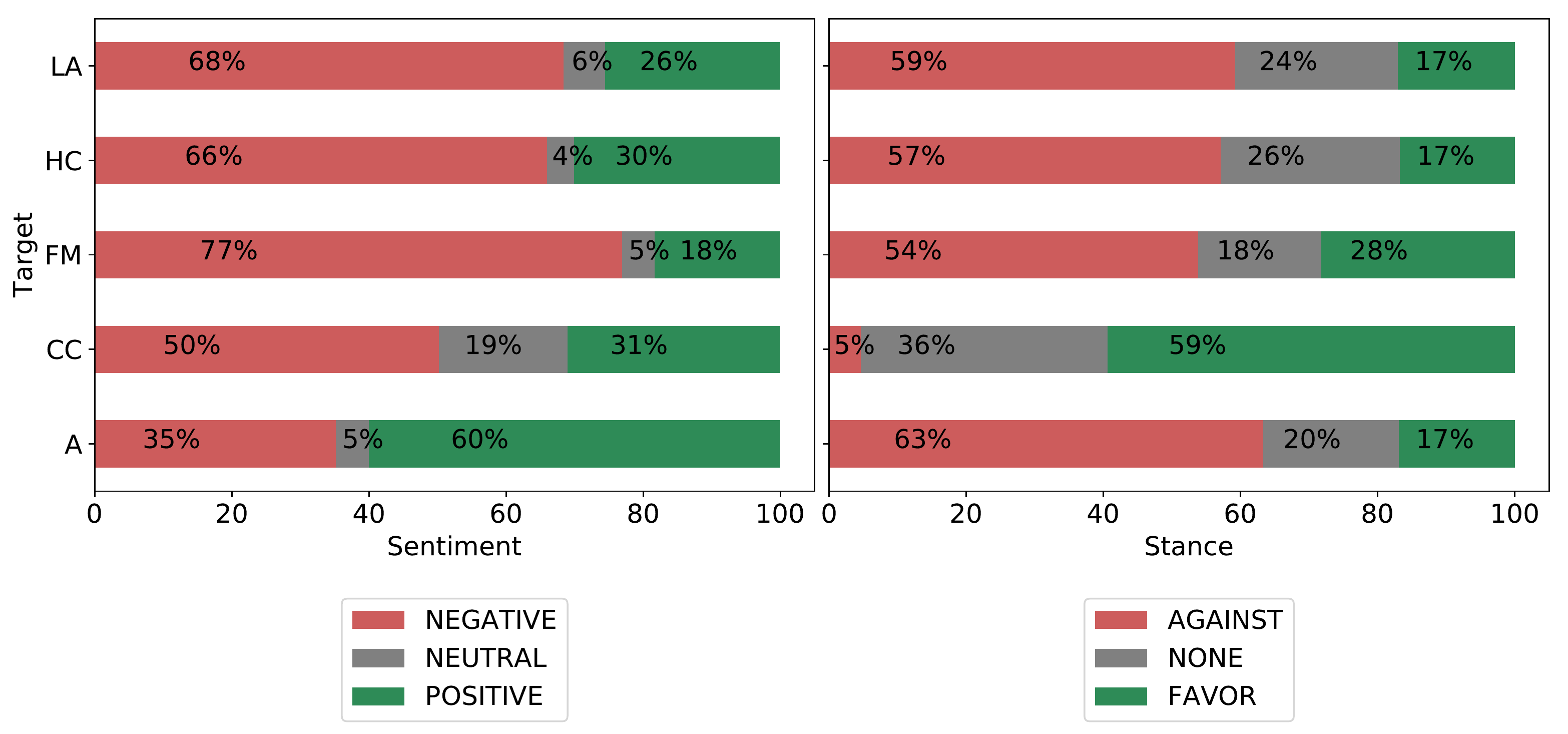}
    \caption{Sentiment and stance distribution}
    \label{fig:Sent_stance_distribution}
  \end{minipage}
\end{figure}

In order to show the overall agreement between sentiment and stance, we use a SemEval stance detection dataset, wherein the tweets are annotated with sentiment and stance labels. The dataset contains a set of 4,063 tweets covering five topics. These topics are 'Atheism' (A), 'Climate change is a real concern' (CC), 'Feminist movement' (FM), 'Hillary Clinton' (HC), and 'Legalization of abortion' (LA).
Fig~\ref{fig:heat_map} illustrates the sentiment distribution on stances for the whole collection (favor, against, and none). The graph shows that the negative sentiment constitutes the major polarity over the three stance categories, as the negative sentiment represents over 56\% of the different stances. This reveals the tendency of using negative sentiments to express a viewpoint on a controversial topic.  

The overall agreement between stance and sentiment is relatively minuscule. About 35\% of  the "favor" stance tweets have a positive sentiment. It can be observed that tweets with neutral sentiments fail to infer neutral stances, with only 12.4\% of the tweets showing neither a positive nor negative polarity. Instead,  54.8\% of the tweets have a negative polarity, while 32.9\% of them carry a positive sentiment for an "against" stance. Thus, it is evident that sentiment does not simply represent stance. 

A large amount of work uses sentiment as a proxy to interpret stances. Based on the assumption that the polarity of a text reveals the stance of the text's author, it has been shown that a stance could be inferred independently of the author's emotional state \cite{li2019multi,sobhani2016detecting,mohammad_stance_2017,mohammad_challenges_2017}. 
A recent study by \cite{li2019multi} used sentiment to predict the stance through a multi-task learning model. In their study, the researchers used a sentiment lexicon along with to build a stance lexicon and incorporated it for the attention layer in the model. Another work by \cite{chauhan_attention_2019} used sentiment as an auxiliary task to predict the stance using a SemEval stance dataset.

It is not necessarily accurate to use sentiment polarity to analyze social media as a means to interpret the general stance toward a topic.  Fig~\ref{fig:Sent_stance_distribution} shows the clear differences in detecting the users' attitudes toward five SemEval topics. While  60\% of the tweets show a positive sentiment toward atheism, the majority (63\%) is opposed to this topic. Similarly, in analyzing the general attitude toward climate change, 50\% of the tweets have negative sentiments, while 59\% of the tweets actually support the idea that climate change is a real concern.  Following the same trend, pure sentiment value fails to detect the accurate stance toward Hillary Clinton, the feminist movement, and the legalization of abortion,  as these distributions show about 9\% of a difference between the accurate "against" position and negative sentiment value for Hillary Clinton and the legalization of Abortion. In the feminist movement, the difference is laudable with a 23\% misinterpretation of the real stance.

Further, we analyzed whether sentiment and stance are independent of each other. We used Cramer's test \cite{cramer2016mathematical} to gauge the strength of relationship between sentiment and stance. The result from Cramer's test indicate that the variance (V) value ranges between 0 and 1, with 1 as an indication of a high association between the nominal variables \cite{liebetrau1983measures}. By applying Cramer's test in the SemEval stance dataset, the resultant V value = 0.12, which is a strong indication of the independence of sentiment and stance from each other.

\section{Expressed vs Unexpressed Stance Detection}
\label{sec:Expressed}
We can divide the literature on stance detection using social media data according to the status of the stance to be modeled, either as expressed in text or unexpressed.
The first type of work is mainly referred to as stance classification, and it represents the majority of work in the exisiting literature. The other type of work is concerned with detecting the stance prior to an event, which is called stance prediction. 

\subsection{Stance Detection for Expressed Views}

Many social media users express their support or opposition toward various topics online, which makes these platforms valuable sources to analyze public opinions toward these topics. This has motivated a rich body of research to focus on inferring the users' stances from their posts expressing their position.

The underlying method for this type is based on using a predefined set of keywords concerning the target of analysis.  Most of the available stance datasets are concerned with studying the users' stance toward a post-event. For instance, a SemEval stance dataset \cite{mohammad2016} created three sets of hashtags for each event or entity to collect tweets concerning three stances. For example, for the topic "Hilary Clinton," the dataset used favor hashtags, such as \#Hillary4President, against hashtags, such as \#HillNo, and ambiguous hashtags. Ambiguous hashtags contain the target without a direct indication of the stance in the hashtag (e.g., \#Hillary2016). In a study done by \cite{darwish2017trump} to predict public opinion and what went viral during the 2016 US elections, tweets that were filtered out by using a set of 38 keywords related to the US elections for streaming relevant tweets to this event were used. Another study employed  tweets generated by users to infer the opinion toward the independence of Catalonia and collected tweets by using two keywords related to the hashtags \#Independencia and \#27S  \cite{taule2017overview}. Similarly, to study mass shooting events, a study by \cite{demszky2019analyzing} made use of Gun Violence Archive and collected list of events between 2015 and 2018 to define a set of keywords that could be used to retrieve the tweets. Further, the study used the Twitter Firehose \footnote{\url{https://developer.twitter.com/en/docs/tweets/compliance/api-reference/compliance-firehose}} to collect tweets for these keywords. 

After collecting the event-related data, it was annotated using predefined guidelines for labelling the data with the stance toward the given target; this method is applied for labelling the SemEval dataset \cite{mohammad_stance_2017,mohammad2016}.

Most of the work on detecting the stance in the expressed views dataset applies NLP methods that model the task as text entailment to detect whether a piece of text is in favor of or against the target \cite{siddiqua_tweet_2019-1,sobhani2019exploring,simaki2017stance,siddiqua2018stance,augenstein_stance_2016,liu_iucl_2016,dias_inf-ufrgs-opinion-mining_2016,igarashi_tohoku_2016}. For instance, the work of \cite{siddiqua_tweet_2019-1} used a SemEval stance dataset which consisted of tweets and target entities to build a textual entailment model that predicted  a target-specific stance. However, some studies showed that employing additional signals from the user's network can further improve the stance detection performance, as will be discussed later in section \ref{sec:network_features}.

\subsection{Predicting the Unexpressed Views}

 Stance prediction aims to infer the stances of social media users with no explicit or implicit expression of these stances online. It is also used  to predict a user's view on events that have not occurred yet.

In stance predication, most studies tend to predict the users' stances on two levels, that is, micro and macro. At the micro level, stance prediction means the estimation of an individual user's standpoint toward a target or event in advance (pre-event), indicating the likelihood that the user will be in favour of or against the target event. This methodology is similar to recommending new items based on a user's previous history of purchases.
At the macro level, the public opinion toward an event is inferred, and the research tend to address this level of prediction as an aggregation of micro predictions\cite{qiu_modeling_2015}. Similarly, \cite{rui_weakly-guided_2017} designed a micro-level stance detection model based on users' previous posts along with a logistic regression to predict stances for new users that were not included in the dataset. Their model produced word distributions for the against/support stances on the topic by using rules derived from user's interactions. Another study \cite{gu_topic-factorized_2014} used matrix factorization to predict micro-level stances based on the voting patterns of the users and the topic model. The study by \cite{gu_ideology_2016} predicted ideology using heterogeneous links between the users. 

For instance, \cite{darwish_improved_2017} investigated the best social media features to predict user attitudes toward new events\cite{darwish_improved_2017}. This work differed from the SemEval-2016 task 6 (a target-specific task) in that the stance of the users was predicted based on each user's historical collection of tweets, instead of a target-relevant single tweet. Rather than focusing on the tweet content (textual features), \cite{darwish_improved_2017} used the similarity between users and inferred latent group beliefs as features in their model, and further employed a bipartite graph along with graph reinforcement. \cite{lahoti2018joint} followed the same concept as \cite{darwish_improved_2017} to define network features, and used a joint matrix in which users and source contents were represented in the same latent space to balance message distribution across social media. A different method was proposed by  \cite{gottipati2013predicting} to predict an individual's unexpressed stance by implementing a collaborative filtering approach and constructing a user ideology matrix that represented a user's ideological stance for an event. 

In macro-level stance prediction, few studies have addressed stance prediction. Most of these studies were customized to analyze specific cases, such as Islamophobia \cite{magdy__2016,darwish_predicting_2017}. Besides analyzing events, these studies carried out a further complementary study for the prediction of  user's stances in the aftermath of an event, based on previous tweets and user interaction. In debate forums,\cite{qiu_modeling_2015} have proposed a micro-level stance prediction model based on user behaviour toward new events in which they have not participated. In this study, this kind of user was referred to as a cold-start user, which is a well-known term commonly used in recommendation systems. In addition, they introduced a macro-level stance prediction model as an aggregation of each user's stance (i.e., at the micro-level). 

\section{Stance detection according to target}
As discussed earlier, the stance detection task needs the presence of a defined target to detect the stance toward it. In the existing literature, stance detection can be be categorized according to the type of the target of evaluation into three groups, namely, single defined targets, multi-related-targets, and claim-based targets. In the following sections, a further explanation of each type is provided.  

\subsection {Target-Specific Stance Detection}
The basic form of stance detection on social media is the \textit{target-specific} stance detection. Most of the previous studies focused on inferring the stance for a set of predefined targets \cite{mohammad2016,kareem_2019,augenstein_stance_2016,liu_iucl_2016,dias_inf-ufrgs-opinion-mining_2016,igarashi_tohoku_2016}. In this type of stance detection, the text (T) or the user (U) is the main input to predict the stance toward a specific predefined single target (G). 

\begin{equation} \label{eq:Stance_target1}
\begin{split}
Stance(T|U,G) = \{Favor, Against, None\}
\end{split}
\end{equation}

The above equation means that a separate stance classification model must be built for each target (G) separately, unlike that in the case of sentiment, wherein a general model can be trained for target-independent sentiment analysis. This is the basic practice, even for benchmark datasets such as SemEval 2016, which covers multiple topics. Most of the published work on this dataset has trained a separate model for each topic (target) separately~\cite{mohammad2016,aldayel_your_2019,siddiqua2018stance}. However, there have been recent trials to apply the transfer learning model between different targets, as will be discussed in more detail in section \ref{sec:transfer}.

\subsection{Multi-related-target stance detection}

In multi-target stance detection, the goal is to jointly learn the social media users' orientation toward two or more targets for a single topic\cite{sobhani_dataset_2017}. The main assumption behind this kind of stance detection is that when a person gives their stance for one target, it provides information on their stance toward other related targets.  
\begin{equation} \label{eq:Stance_multi-target}
Stance(T|U,G_n) = \{(Favour G_1, Against  G_{n+1}),(Favour of G_{n+1}, Against G_1)\}
\end{equation}

For instance, a tweet could express a stance toward multiple US presidential candidates at the same time, so when users express their in-favor stance toward Hillary Clinton, it implies an against stance toward Trump \cite{darwish2017trump,lai2016friends}. In ~\cite{sobhani_dataset_2017} work, the first multi-target stance detection data-set is presented, containing 4,455 tweets related to the 2016 US elections. In order to detect the subjectivity toward two targets, \cite{sobhani_dataset_2017} used an attention-based bidirectional recurrent neural network (RNN) to jointly learn the stances toward two related targets. The notion of multi-target stance detection has been usually used to analyze the relation between two political candidates, such as Clinton and Trump, by using domain knowledge about these targets to improve the classification performance \cite{lai2016friends}. Following the same approach, the study by \cite{darwish2017trump} constructed a dataset with 3,450 tweets annotated with stance labels for the two US 2016 election candidates (Trump and Clinton) at the same time. Furthermore, the work of \cite{wei2018multi} proposed a memory-based algorithm focusing on jointly modeling multiple targets at the same time. Their memory-based model provides the current state-of-the-art result so far on the multi-target benchmark dataset.

\subsection{Claim-based stance detection}
\label{sec:claim_based}
In claim-based or open-domain stance detection, the target of the analysis is not an explicit entity, as is the case in the ones discussed earlier. This, however, is a claim in a piece of news. The first stage to detect a stance is to identify the main target claim from the sequence of conversation or given text. The main input to the claim-based stance detection model is the claim (C), which could be the rumor's post or based on an article headline. In the fake news task, the claim tends to be the article headline and the text is the article body. On the other hand, for the rumor's veracity task, the main input to be evaluated is the rumor's post, while the text is the replies to the rumors. The prediction label sets tend to take the form of confirming the claim or denying it. 
\begin{equation} \label{eq:Stance_target2}
\begin{split}
Stance(T,C) = \{Confirming, Denying, Observing\}
\end{split}
\end{equation}

Claim-based stance detection is considered a suitable method to analyze the veracity of the news. For that reason, claim-based stance detection has been heavily used for rumor resolution studies \cite{hamidian2015rumor,aker_simple_2017,zubiaga_discourse-aware_2018,gorrell-etal-2019-semeval}. In a study by \cite{hamidian2015rumor}, a supervised learning model was used, along with a new set of features known as "pragmatic features" which contained the named entity, event, sentiment, and emoticons. Interestingly, \cite{aker_simple_2017} concluded that problem-specific feature engineering outperformed the other state-of-the-art systems in rumor identification tasks. Their model, which used a random forest, outperformed the advanced LSTM-based sequential models in SemEval-2017 task 8 proposed by \cite{kochkina_turing_2017}. Along same lines, the study of \cite{aker_simple_2017} used the same feature engineering approach proposed by \cite{elfardy2016cu}, in which lexical and semantic features were used to help with identifying the stance of a Twitter user. More recently, for the conversation-based tasks, the study by \cite{zubiaga_discourse-aware_2018} showed that LSTM can outperform other sequential classifiers and feature-based models. 

In another study by \cite{li2019rumor}, multi-task learning was used with a stance detection layer to classify the stance of a given tweet as supporting or denying a given claim.

\section{Stance modeling on social media}

\begin{figure}
\centering
\begin{minipage}[b]{0.99\textwidth}
    \includegraphics[width=\textwidth]{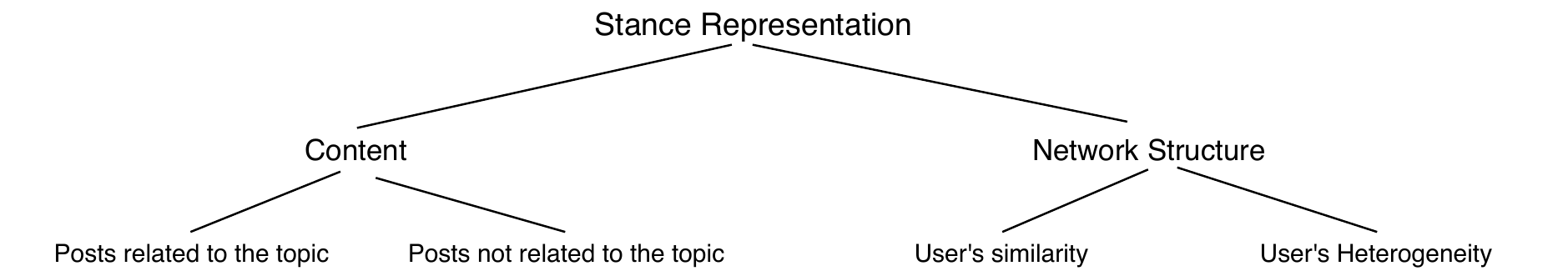}
    \caption{Stance representation in social media}
    \label{fig:Stance representation}
  \end{minipage}
\end{figure}

Stance on social media has been modeled using various online signals, which have been used as features for training the stance detection models. These signals can be categorized into two main types, namely, 1) content signals, such as the text of the tweets, and 2) network signals, such as the users' connections and interactions on their social networks. Figure \ref{fig:Stance representation} summarizes the main features used in each category to model stances on social media. In the following section, we describe each of these two categories of signals and how they have been used for detecting stances. Finally, we present a comparison of their effectiveness for detecting stances on multiple datasets.

\subsection{Content Features}
This section discusses stance representation, focusing on the textual features derived from the users' content online. As illustrated in Figure \ref{fig:Stance representation}, the content can be collected based on the topic of the analysis. In this kind of feature representation, data is collected based on the range of keywords reflecting the topic. Another type of representation is concerned with collecting content that has no direct relation to the topic. The main objective in this case is to model the stance based on a user's behavioral data, rather than topic-level stance detection.

Furthermore, the content features can be categorized into two general types, namely, the linguistic features and users' vocabulary. The first type of features are related to the text's linguistic features that help in inferring the stance. The other type concerns modeling a user's stance based on the user's choice of vocabulary. 

\subsubsection{Linguistics Features}
The majority of work on stance detection has been focused on utilizing the linguistic elements that capture a social media user's stance. This massive dependency on linguistic cues is due to definition of stance detection as a textual entailment task, wherein the task is to detect a stance in a given piece of text (e.g., tweet~\cite{mohammad2016}). In the existing literature, the stance detection work that is concerned with using textual cues to detect stances includes textual features, sentiment polarity, and latent semantics. 

For instance, using textual features such as the n-gram modeling of the text has been heavily investigated in the existing literature \cite{anand_cats_2011,mohammad_stance_2017,sobhani2017stance}. Using the n-gram modeling of the text shows promising results. In the SemEval-2016 stance detection task, the use of word and character n-gram modeling obtained the best f-score among the other participating systems \cite{mohammad2016}. Another textual feature that has been used to infer the stance is the sentiment polarity of the post \cite{mohammad_challenges_2017,elfardy2016cu,elfardy2017perspective}. In general, the use of sentiment as a feature was not sufficient in predicting the stance, as some studies have concluded \cite{mohammad_stance_2017,elfardy2017perspective}. 
Another kind of feature is the latent semantics feature that aims to reduce the dimension of a given input, such as mapping the sentences according to a predefined set of topics (i.e., topic modeling) \cite{elfardy2016cu}. Topic modeling has been applied by several different studies \cite{elfardy2016cu}. For instance, \cite{elfardy2016cu} used weighted textual matrix factorization  (WTMF) and frame-semantic parsing to model a given tweet that had been used as feature by \cite{patra2016ju_nlp} to map unigrams to the topic sphere. Another study used a simple bag-of-topics to model target words. 

\subsubsection{Users' Vocabulary Choice}
There are a considerable number of studies that represent stance detection based on the users' vocabulary. The hypothesis behind using this kind of user modeling is that individuals with same stance tend to use the same vocabulary choices to express their points of view \cite{darwish2019unsupervised}. 

The focus of these studies is mainly to disentangle the topic from the viewpoint where the vocabulary is not only linked to the topic but also individual attitudes and characteristics \cite{klebanov2010vocabulary}. For instance, people with an against stance to abortion tend to use terms such as "pro-life" to express their opposing stance. The work of \cite{klebanov2010vocabulary} built a user stance detection model based on users' vocabulary choices through a generative model and discriminative model using Naive Bayes and SVM classifiers, respectively. Another work, \cite{rui_weakly-guided_2017} proposed a word-generative model based on the users' interaction to build a set of word representations for each stance toward a topic. In addition, the work of \cite{benton2018using} followed the same direction, wherein the users' interactions helped in shaping the set of vocabulary used to identify the stance. Another study by ~\cite{zhu2019hierarchical} used a hierarchical topic for opinion discovery, based on the author of the text vocabulary. Lately, the work on vocabulary choice has been linked with the user behavior on social media to further improve the stance prediction model. The work of \cite{rui_weakly-guided_2017} used a generative model based on the users' keyword choices for each standpoint, along with their interactions as regularization to predict the stance.  In addition \cite{li2018structured}, introduced user interactions and post embedding by using an embedding vector for the pro-stance and a different vector for the con-stance.

\subsection{Network Features}
\label{sec:network_features}
Social media provides a special kind of social data due to the structure of its platforms, wherein users can be characterized and/or analyzed based on their social connections and interactions. 

Many existing works used network features to gauge the similarity between the users.

The network features that have been used to learn the users' representations on social media can be grouped under two categories, namely, users' behavioral data \cite{thonet2017users,darwish_improved_2017,kareem_2019} and users' meta-data attributes \cite{pennacchiotti2011democrats}. The application of users' behavioral data to identify the stance is motivated by the notion of homophily, based on the social phenomenon according to which "individuals associate with similar ones" ~\cite{bessi2016homophily}. When it comes to social media, the similarity between users is considered a core property that helps in inferring stances.

The interaction elements have been used to define the similarity between the users. One of the elements that has been extensively used to infer Twitter users' stances is the retweet \cite{borge2015content,darwish_predicting_2017,weber2013secular,rajadesingan2014identifying}. Another element that has been heavily investigated is the hashtag; this element has been used in literature to infer similarity between users to predict the stance \cite{darwish_improved_2017,dey2017assessing}.  The work of \cite{dey2017assessing} used the soft cosine similarity feature to gauge the similarity between the users that post under the same hashtag. The work of \cite{darwish_improved_2017} used graph reinforcement to calculate the similarity between the users that post under the same hashtag.  

In a recent study by \cite{aldayel_your_2019}, three types of network features to  model stances on social media were defined. These network features are the 1) interaction network, 2) preferences network, and 3) connection network. The interaction network represents the users' direct interactions with other users, in the sense of retweets, mentions, and replies. This type of network provides the best performance score of the stance detection model in compression with two other networks. The preference network is the network of others users that post or are mentioned in the tweets the users like. This network allows the detection of stances  for users who may have limited posting or interaction behaviors online. Finally, the connection network includes the friends and followers of the users. The three types of networks provide the best performance in comparison with content features \ref{table:comparitive_analysis}. 

Another study by \cite{fraisier2018stance} introduced a multi-layer graph model to represent profiles from different platforms and extract communities. By doing so, the model allowed a stance to diffuse from the set of known profiles and predict profiles' stances by using  pairwise similarities between the users' profiles.  
 
The use of network features has been beneficial in detecting social media users' stances and future attitudes in the aftermath of an event (i.e., stance prediction). For instance, in a study done by \cite{darwish_predicting_2017}, a user's similarity components was utilizied as the features. In this work, the similarity was calculated based on the interaction elements of a given tweet. These interaction elements included mentions, retweets,  and  replies, as well as  sebsite links (URLs) and hashtags used by users in their tweets. Similarly, the work of \cite{thonet2017users} used the retweet and reply features to define the users' social networks.

 Another line of study has used heterophily to measure the dissimilarity between the users to model the stance \cite{trabelsi2018unsupervised}. For instance, the study by \cite{trabelsi2018unsupervised} was based on the tendency of a user to reply to the opposed viewpoint. The study used a rebuttal variable to model the users' interactions and denote if the replies attacked the previous author's parent post. The value of the rebuttal depended on the degree of opposition between the viewpoint of the parent post and parent tweet.

\subsection{ Comparative Analysis of Stance Modeling } 
\label{sec:analysis_NW_txt}
Table \ref{table:comparitive_analysis} shows a comparative analysis of the stance-detection performance using the two types of modeling, content versus network. In addition, it reports the results of studies that used a combination of both these types. It must be noted that using network features to model the stance outperforms the content modeling of the stance. Textual modeling has the lowest performance as this type of modeling depends on the textual entailment task, while the network features provide the highest performance in comparison with the former. This may explain the fact that using network features puts the bigger picture of modeling the users' attitudes using their interactions and similarities into consideration. Using network features overcomes the limitations posed by the textual entailment modeling of stance. As demonstrated by \cite{aldayel_your_2019}'s this study used a SemEval dataset to model stances on social media using the model with the best reporting results on stance datasets to compare the performance of the stance, using content and network features. This study provides a thorough comparative analysis of the stance's overall performance textual and network features are used. The use of network features provides the best performance in compression with previous studies, wherein textual modeling along with transfer learning methods were used, as reported by \cite{mohammad2016}. Furthermore, a study by \cite{lahoti2018joint} showed that the network features outperform the textual modeling of stance when using non-negative matrix factorization for stance detection. The study by \cite{darwish_improved_2017} confirmed the same conclusion, wherein the use of network features outperformed the textual representation using two datasets (Islam and Island datasets). As the study by \cite{magdy__2016} indicated, the network modeling of stance outperformed the use of textual modeling for the expressed and non-expressed stances on social media.

\begin{table} [t!]
\caption{ A comparison of stance detection models using network , content and both as features. }{%
	\begin{tabularx}{\textwidth}{|>{\leavevmode}p{3.6cm}|>{\leavevmode} p{1cm}| >{\leavevmode} p{1cm}| >{\leavevmode} p{1cm}| >{\leavevmode} p{3.5cm}| >{\leavevmode} X|}
		\hline
\textbf{Dataset} & \textbf{NW} & \textbf{Content} & \textbf{Both} & \textbf{ML Algorithm} & \textbf{Study} \\ 
     \hline
 Before Paris attacks  &\textbf{85.0} & 82.0 & 84.0 & SVM  & \cite{magdy__2016} \\
 Islam Datase &\textbf{84.0} & 76.0 & -& SVM & \cite{darwish_improved_2017} \\
 
 Islands Datase & \textbf{79.0} & 71.0 & - & SVM & \\
 Gun control, abortion and Obama care &  \textbf{81.9} & 60.6  & 82.0 & Non-negative matrix factorization &\cite{lahoti2018joint} \\
 
 SemEval stance dataset & 71.6 & 69.8 & \textbf{72.5} & SVM & \cite{aldayel_your_2019} \\
 SemEval stance dataset & 61.8 & 62.8 &\textbf{65.9} & SVM & \cite{lynn2019tweet}\\
  
  LIAC dataset & \textbf{67.8} & 54.1 & - & Generative Model & \cite{rui_weakly-guided_2017} \\
 
 \hline
\end{tabularx}}
\label{table:comparitive_analysis}
\end{table}

\section{Stance Detection Algorithms}

In this section, the main machine learning (ML) algorithms used for stance detection are discussed. According to the existing literature, the ML algorithms used for stance detection on social media can be divided into three main approaches, namely, 1) supervised learning, 2) weakly-supervised and transfer-learning, and 3) unsupervised stance detection. In the following sections, each of these approaches are discussed in greater detail. 

\subsection{Supervised Learning}

Supervised learning is the basic and most common approach for most of the work on stance detection  \cite{zhang_stances_2019,lai_multilingual_2020,walker2012your,krejzl_uwb_2016,igarashi_tohoku_2016,gottipati2013predicting}. In this approach, a stance dataset is annotated using a predefined set of labels, usually two (pro/against) or three (pro/against/none) labels. For example, the SemEval-2016 stance dataset uses three labels, "In-Favor," "Against," and "None" \cite{mohammad2016} for a set of five different topics. Many studies have been published on this dataset using different supervised ML algorithms such as classical algorithms, for instance, Naive Bayes (NB), SVM, and decision trees, and deep learning algorithms, for example, RNNs, and LSTMs, to detect the stance on this labelled dataset.   

For example, the work of \cite{mohammad_stance_2017} used an SVM model with linguistic and sentiment features to predict the stance. This study showed that the use of content features only (i.e., n-gram) provided an F1 score equal to 69.0\%, which surpassed the use of sentiment as a feature with an F1 score of about 66.8\%. Another study by \cite{walker2012your} used labelled data toward 14 topics to train an NB model to predict the stance.The work of \cite{elfardy2016cu} used an SVM model as well as lexical and semantic features to classify the stance in SemEval,  which had an F1 score equal to 63.6\%. Another study by \cite{wojatzki2016ltl} used a stacked classifier and syntactic features to classify the stance in a SemEval stance dataset. Their model showed a minuscule improvement on the overall stance detection performance with a 62\% F1 score. The work of \cite{siddiqua_tweet_2019-1} proposed a neural ensemble model using bidirectional LSTM on a SemEval stance dataset along with a fast-text embedding layer. Their model indicated an improvement on the overall F score, raising it to 72.1\%. A recent work by \cite{li2019multi} used a bidirectional gated recurrent unit to build a multitask learning model that leveraged the sentiment of a tweet to detect the stance. This model showed an improvement of the stance detection model's overall F score of the SemEval dataset to reach a score equal to 72.3\%. A detailed comparison of each study's performance on the SemEval dataset is reported later in section \ref{sec:best_algorithm}.

 
\subsection{Transfer Learning}
 \label{sec:transfer}

To address the scarcity of the labelled data for each target in the stance detection task, some studies in this field attempted to incorporate transfer learning techniques to enrich the representation of targets in the dataset and enhance the overall performance of stance detection  \cite{dias_inf-ufrgs-opinion-mining_2016,augenstein2016usfd}.
In transfer learning, the knowledge that an algorithm has learned from one task is applied to a separate task, such that in the task in stance detection, transfer learning is applied across different targets. One of the well known stance detection datasets that motivated the work on transfer learning is the SemEval stance (Task B) \cite{mohammad2016}. This dataset contains 78,000 unlabelled tweets related to Trump and provides a good source for research, to be explored through  various transfer-learning algorithms.  For example, in the work of \cite{augenstein2016usfd}, the SemEval stance (Task B) was used to train a bag-of-word auto-encoder along with Hillary-labelled data to detect the stance. Another study by \cite{dias_inf-ufrgs-opinion-mining_2016} used Hillary Clinton's labelled data along with Trump's unlabelled data to develop a rule-based algorithm to help in detecting the stance for the SemEval stance Task B. In the work of \cite{wei_pkudblab_2016}, a CNN model was used with Google news embedding on the SemEval stance Task B. For Subtask B, their model trained on two class datasets and further, by using a modified softmax layer, the classification of three classes with a voting scheme was performed. With these configurations, their model ranked among the top three models with an overall F-measure of 56.28 for Task B. Moreover, the study by \cite{zarrella_mitre_2016} implemented distant supervision using SemEval ( Task B) to learn the features from two unlabelled datasets by using hashtag prediction. Another study by \cite{wei_modeling_2019} used topic transfer knowledge to leverage the shared topics between two targets and predict the stance for unlabelled data. Recently, many studies have incorporated the concept of transfer learning using new datasets other than the SemEval stance (Task B). A recent study by \cite{hanawa2019stance} used Wikipedia articles to build knowledge extraction for each topic on a dataset that contained seven topics from Wikipedia. Another work by \cite{ferreira2019incorporating} used three datasets, namely blogs,  US  elections, and  Moral Foundations Twitter, and designed algorithms that incorporated label dependencies between them to train the stance detection model.    Another study by \cite{sen2020reliability} used weak supervised stance detection to predict the stance towards 7 political targets to analyze the  politicians' approval from tweets. In their model they used distance supervision method (DSSD) to label the data using certain keywords and hashtags. They found that using targeted oriented method (DSSD) is not generalized for unseen targets,such as Macron and Putin. One reason behind that is due to the heuristics used to generate weakly labeled data to train (DSSD) may not hold in different time periods. They concluded that low resource methods, stance detection trained on topic specific data,is more robust in predicting the stance in comparison with off the shelf methods such as using sentiment analysis VADER.   

\subsection{Unsupervised Learning}
Recently, attention has been devoted toward building unsupervised stance detection models. In these kinds of studies,  clustering techniques are primarily used with a focus on the user and topic representation on the social media platform \cite{darwish2019unsupervised,joshi2016political,trabelsi2018unsupervised,conforti2021will2}. The work of \cite{trabelsi2018unsupervised} proposed an unsupervised model using the clustering model at the author and topic levels. In this study, six topics collected from two online debate forums, namely, 4Forums and CreateDebate were used. Their clustering model leveraged the content and interaction networks of the users (i.e., retweets and replies).

A recent study by \cite{darwish2019unsupervised} used the clustering technique to create an initial set of stance partition for annotation. In their work, the researchers used unlabelled tweets related to three topics, Kavanaugh, Trump, and Erdogan. Their findings showed that using retweets as a feature provided the best performance score upon implementing clustering algorithm (DBSCAN), which surpassed the supervised method when using the fast-text and SVM models. Their findings are considered greatly motivational for the use unsupervised methods for stance classification in the future. 

\subsection{Best Performing Algorithms}
\label{sec:best_algorithm}

\begin{table} [t!]
\caption{ Comparing the stance detection models on SemEval stance dataset. }{%
	\begin{tabularx}{\textwidth}{|>{\leavevmode}p{1.5cm}|>{\leavevmode} p{4.3cm}| >{\leavevmode} p{4.2cm}| >{\leavevmode} l| >{\leavevmode} X|} 
		
		\hline
\textbf{Algorithm} & \textbf{Model} & \textbf{Features} & \textbf{F1}  & \textbf{Study} \\ 
	\hline
       & SVM & NW(Mentions+urls) & 71.56 &  \cite{aldayel_your_2019} \\
Supervised & SVM & NW(Mentions+urls)+content & \textbf{72.5} &  \cite{aldayel_your_2019} \\
learning &  RNN & NW (followee) & 61.8 & \cite{lynn2019tweet} \\ 
  &  RNN & NW (followee)+Content & 65.9 & \cite{lynn2019tweet} \\ 
  &   Nested  LSTMs & Content & 72.1 & \cite{siddiqua_tweet_2019-1} \\
 & bidirectional gated recurrent unit & Content & 72.33 & \cite{li2019multi} \\
  & Hierarchical Attention NN & Content & 69.8 & \cite{sun2018stance} \\
  &  SVM & Content & 70.0 & \cite{siddiqua2018stance} \\

\hline
Transfer & BiGRU & Noisy labeling + topic modeling& 60.8 & \cite{wei2019topic} \\
learning & BiLSTM & Sentiment lexicon & \textbf{65.3} & \cite{li-caragea-2019-multi}\\
 & Deep averaging network & words embedding & 35.2 & \cite{ebner2019bag}\\
  & Deep averaging network & Glove & 30.2 & \cite{ebner2019bag}\\
 \hline
\end{tabularx}}
\label{table:comparitive_analysis_algorithms}
\end{table}

Since the SemEval stance dataset was the most-used dataset for bench-marking the performance on stance detection by using multiple ML approaches, this dataset has been used in this section to compare different ML approaches discussed in the above sections.
Table \ref{table:comparitive_analysis_algorithms} shows the best performing models based on the type of algorithm.  As is expected, the transfer learning models have a lower performance score compared to supervised learning models. However, the study by \cite{li-caragea-2019-multi} shows promising performance, with an F-score of 0.653. Nevertheless, other trials, such as the work of \cite{ebner2019bag}, which used a deep averaging network (DAN) with GloVe word embedding, achieved an average F1 score of around 0.3\%, which is close to random performance.

As is evident in Table \ref{table:comparitive_analysis_algorithms}, the supervised algorithms are more effective for stance detection when they achieve higher F-scores on the SemEval dataset than transfer-learning approaches. This can be seen in the models that have incorporated network features to detect the stance. It is interesting to see that simpler machine learning models, such as SVM are more effective than deep learning models. In addition, these models achieve even better performances when they incorporate network features along with content, such that using both with a simple linear SVM is more effective than using word embedding with RNNs and LSTMs. 

There are no unsupervised algorithms that have been applied to the same SemEval dataset to date. The only reported results, as discussed earlier, are those by \cite{kareem_2019},  wherein the model was applied on different datasets (Trump, Kavanaugh, and Erdogan datasets). The unsupervised model of this study used network features and outperformed the model with the use of fast-text word embedding with a clustering purity equal to 98\%. Overall, it can be observed that the best performing models are the ones framed in user and social contexts with the use of user network features in the stance detection task.

\vspace{0.5cm}

\section{Stance Detection Applications}

Stance detection has been mainly used to identify the attitude toward an entity of analysis to allow the measurement of public opinion toward an event or entity. However, there are other applications for stance detection that are discussed in this section.

\subsection{Analytical Studies}
The use of stance detection has proven beneficial as a social sensing technique to measure public support related to social, religious, and political topics.

Some examples of the use of stance detection for analyzing political topics are the studies on analyzing public reaction toward Brexit based on Twitter data \cite{lai2020multilingual,simaki2017annotating,grvcar2017stance,simaki2017stance}. Furthermore, most of the stance detection studies used data from the 2016 US elections to analyze the stance toward two candidates, Hillary Clinton and Donald Trump \cite{lai2016friends,darwish2017trump}.  
The work of \cite{lai2018stance} analyzed the stance toward the political debates on Twitter regarding the Italian constitutional referendum held in December 2016. Another work by \cite{taule_overview_2017} studied the public stance toward Catalan independence. A more recent study by \cite{lai2020multilingual} used a multilingual dataset to study political debates on social media. The researchers collected entities and events related to politics in five languages, English, French, Italian, Spanish, and Catalan.  

Another line of study uses stance detection to analyze the public viewpoint on social aspects. The public opinion toward immigration has been recently studied \cite{gualda2016refugee,bartlett2015immigration}. The work of \cite{gualda2016refugee} examined the attitude toward the refugees using data from Twitter. They collected tweets using the keyword "refugees" in different languages. They used the sentiment of the discourse to analyze the public stance toward the refugees. Another work by \cite{bartlett2015immigration} studied immigration in the United Kingdom by annotating the tweets related to immigration with sentiment polarity. They used negative polarity as an indication of the against stance and positive sentiment as an indication of the in-support viewpoint. They found that about 5\% of the users were against immigration, while 23\% were in support of the immigration. It worth mentioning that the last two studies mentioned are seen as sub-optimal in measuring the stance due to their heavy reliance on the sentiment of the text, which has been demonstrated in section \ref{sec:senti_analysis} and previous studies \cite{aldayel_assessing_2019,sobhani2017stance} to be sub-optimal. A recent study by \cite{xi2020understanding} analyzed the users' images posted on Facebook to understand their ideological leaning and how political ideology is conveyed through images. In their work, the researchers used the scores generated by the CNN image classifier model to further explore the features that distinguish the liberals from the conservatives. The most distinct features predicting liberal images are related to economic equality, whereas conservative members tend to use images that have objects related to state and economic power, such as "court"  . \textcolor{black}{Another study by \cite{stefanov2020predicting} used social media users stance to predict Media Bias. They used users stance towards eight topics  along with valence score of top users to predict media media leaning. }

On the other hand, stance detection has been used to analyze the attitude related to disruptive events \cite{demszky2019analyzing,darwish_predicting_2017}. For instance, the work of \cite{darwish_predicting_2017} studied the public attitude toward Muslims after the Paris terror attacks in 2015. In their work, the researchers collected tweets that mentioned keywords related to Islam. Subsequently, they analyzed users' interactions and used the historical tweets to predict their attitudes toward Muslims. The work of \cite{demszky2019analyzing} analyzed the attitudes toward 21 mass shooting events. They first derived a list of mass shooting events between 2015 and 2018 from the Gun Violence Archive. For each event, they defined a list of keywords to collect tweets related to those events. Further, to study the polarization of opinion toward these events, they estimated the party of each user, based on the political accounts they followed. 

\subsection{Applications Related to Social-Based Phenomena }

Stance detection has been used to solve algorithmic issues on social media platforms. These issues are a reflection of social phenomena on these platforms. 
 The most common phenomena that have affected various social media platforms are echo-chambers and homophily \cite{fuchs2017social}. Previous studies have concluded that social media are polarized in nature, which boosts homophilic behavior  \cite{barbera2015tweeting,bessi2016homophily,quattrociocchi2016echo,darwish_improved_2017,garimella2018polarization}. Homophily is the social phenomenon that concerns people's tendency to connect with "like-minded friends." An echo-chamber is the cascade of certain information among groups of people.  This social behavior has been magnified in social media structures wherein certain beliefs are amplified within close circles of communication. Consequently, people are exposed to content in consent with the same opinions that they hold. As a result, this reinforces social media users' view biases and blinds the users from other sides of information. Therefore, stance detection has been used to help in measuring and alleviating the problems that result from the polarization on social media. For example, a study by \cite{garimella2017mary} used stance as the main factor to expose users to contradicting views.  Further, stance detection was used to identify and measure the controversy level on social media platforms  \cite{al2018studying,dori2015automated,jang2018explaining}. In a study by \cite{jang2018explaining} the stance summarization technique was used to re-rank controversial topics. This method collected arguments based on five topics and summarized the overall stance with respect to the top tweets for a given topic. 

\subsection{Veracity Checking Applications}
The rapid dissemination of news on social media platforms encourages people to depend on these platforms as their main sources of information. This kind of information consumption triggers critical issues in social media related to the credibility of the exchanged information, namely fake news and rumor detection. Recently, a lot of attention has been devoted toward the classification of stances to help in taking the first step toward solving the veracity checking issue \cite{gorrell-etal-2019-semeval,derczynski_SemEval-2017_2017,allcott_social_2017}. 

A rumor is defined as an uncertain piece of information that lacks a secure standard of evidence for determining whether it is true or false \cite{Levinson1996}. It has been shown that false rumors have obvious effects across various fields. For example, in the healthcare sphere, false rumors on the Web pose major public health concerns. In 2014, rumors about the Ebola epidemic in West Africa emerged on social media, which made it more difficult for healthcare workers to combat the outbreak \cite{shultz20152014}. Rumors also affect modern-day journalism, which depends on social media as the main platform for breaking news. Additionally, there have been numerous occasions where false rumors have yielded severe stock market consequences \cite{schmidt2015stock}. To this end, many initiatives have been introduced, such as the Emergent dataset, PHEME dataset, and RumourEval SemEval-2017 and -2019 datasets \cite{ferreira_emergent:_2016,derczynski_SemEval-2017_2017,kochkina_turing_2017,gorrell-etal-2019-semeval}. These initiatives  have been implemented to encourage the development of tools that can help with the verification of the misinformation in news articles.

For these task, stance detection has been used as a key feature for checking the credibility of a piece of news. As discussed in section \ref{sec:claim_based}, stance in comments toward the news is measured to detect if these comments are confirming or denying the news, which, in  turn, is used to detect if the news is a rumor or authentic.

Unlike rumors, fake news aims to create misleading information and it is always false \cite{zubiaga2017detection}. Data veracity refers to the truthfulness, trustworthiness, and accuracy of the content. Social media posts have massive amounts of user-generated content through which fake information can be easily spread. One particularly challenging task is verifying the truthfulness of social media information, as the content by nature tends to be short and limited in context; these characteristics make it difficult to estimate the truthfulness of social media information compared to other information resources, such as news articles. One of the proposed methods to address the fake news is based on detecting the stances toward news organizations. This is due to the fact that understanding the stances of other organizations toward a news topic is helpful in inferring the truthfulness of the news article. The Fake News Challenge initiative (FNC-1) adopted this approach and proposed a stance detection task to estimate the stance of articles toward a given headline (i.e., claim). The best performing system at the FNC was proposed by \cite{fake17}. In their study, they used gradient-boosted  decision  trees and  a convolutional neural network (CNN), along with  many  textual features. Another study by \cite{mohtarami2018automatic} achieved relatively  similar results to the best system with a feature-light memory network model enhanced with LSTM and CNN. A more recent study by \cite{shu2019beyond} used three features extracted from users' interactions, news authors, and the contents of news article to better detect the fake news. In  the study by \cite{ghanem_stance_2018}, a combination of lexical knowledge, word embedding, and n-gram was to detect the stances in two datasets, FNC-1 and Emergent.  The study by \cite{borges_combining_2019} proposed a new text representation that incorporated the first two sentences of the article along with the news headline and the entire article to train a bidirectional RNN model using an FNC-1 dataset. 

\section{Stance Detection Resources}

This section lists the current available resources for stance detection tasks. Table \ref{table:stance_prediction} shows the available datasets with stance annotation data in a chronological order. This table categorises the datasets on a higher level as classification and prediction datasets, as defined in section ~\ref{sec:Expressed}. In classification tasks the datasets further categorized as: target-specific, multi-target and claim-based stance dataset. The Other type of taxonomy is stance prediction datasets, with macro and micro predictions. The current data sources that has been used for stance detection includes: Twitter \cite{mohammad2016} \cite{ferreira2016emergent} \cite{derczynski_SemEval-2017_2017}, Wikipedia\cite{bar2017stance}, International debate education association website \cite{bar2017stance} and various news sites such as: Snopes.com \cite{ferreira2016emergent}. 

\textit{Target-specific datasets}: Table \ref{table:stance_Dataset} shows a list of available datasets for stance detection on social media data. There are two publicly available datasets that contains stance annotations for predefined targets on social media. The first dataset is the SemEval stance detection \cite{mohammad2016} provides two datasets to serve two frameworks: supervised framework (Task A) and weakly supervised framework (Task B). In the supervised framework, the annotation scheme evaluates the author of a tweet stance towards five targets (atheism, Hillary Clinton, climate change, feminist movement and legalization of abortion). In this dataset, the tweet-target pair is annotated for stance and sentiment labels. On the other hand, the weakly supervised framework dataset contains tweets related to one target (Donald Trump). In this dataset the training data contains 78,000 tweets related to (Donald Trump) without stance labels. Several studies proposed stance detection model for SemEval stance dataset. 

Another work by \cite{gautam2019metooma} provides a dataset related to \textit{(Me Too)} movement. This dataset contains around 9000 tweet annotated with stance, hate-speech relevance, stance dialogue act and sarcasm. Another work by \cite{kuccuk2019tweet} introduce a dataset with stance and name entity labels. The dataset contains 1065 tweets annotated with stance labels towards two entities Galatasaray and Fenerbahc. A recent study by \cite{conforti2020will} provides a dataset for stance detection that contains around 51K tweets covering financial domain. \textcolor{black}{ Another study by \cite{vamvas2020x}  provides a dataset  consists  of  German,  French  and Italian  comments from Smartvote. The dataset constricted to evaluate cross-lingual methods for stance detection towards 150 targets.}

\textit{Claim-based datasets}: In this kind of stance detection dataset the object of evaluation is the source of information instead of social actor. The Rumours dataset \cite{qazvinian2011rumor} is a claim-based stance detection dataset designed for Rumours resolutions. This dataset contains 10,417 tweets related to Obama, Airfrance, cellphone, Michelle and plain.  In this dataset the rumour tweet is evaluated against set of other tweets to define the stance of these tweets in supporting or denying the source of the rumour. Additionally, the Emergent dataset \cite{ferreira_emergent:_2016} is a Claim-based stance detection dataset for  Fact checking. This dataset contains rumours from variety of sources such as rumour sites, e.g.snopes.com, and Twitter. Another dataset available for claim detection, is Fake-News dataset. This dataset contains news articles from the Emergent dataset where the news headline being evaluated against set of a body text.

The SemEval 2019 rumours detection dataset by \cite{gorrell-etal-2019-semeval}, enriched the SemEval 2017 (rumours detection task) dataset by adding new data from Reddit and extend the language representation of this dataset to include Russia.

\textit{Multi-related-targets}: The two datasets that have multi-related-targets stance annotations are the Trump vs. Hillary dataset\cite{darwish2017trump} and Multi-targets dataset \cite{sobhani_dataset_2017}. In Trump vs. Hillary dataset, each tweet is stance annotated for the two candidates in the same time such as (supporting Hillary and Against Trump). The same annotation technique has been used in Multi-targets dataset for an extended list of US presidential candidates.   The Multi-target dataset \cite{sobhani_dataset_2017,sobhani2019exploring} contains three pairs of targets Clinton-Sander, Clinton-Trump and Cruz-Trump.

\textit {Stance prediction datasets}:
As a result of the lack of benchmarks in this kind of stance detection, the researchers tend to build their own datasets as illustrated in tables \ref{table:stance_prediction}. These datasets are designed to predict the stance before the event time. There are two datasets constructed by \cite{darwish_improved_2017}, where they used twitter as a source for stance prediction data. 

\begin{table} [hb]
\caption{data-set with stance annotations for prediction task (on Chronological order)}{%
	\begin{tabularx}{\textwidth}{|l|p{1.6cm}|p{1.6cm}|p{4cm}|X|}
		\hline
Data-set & Type & Data Source & Stance definition (annotation) & Size 
\\ \hline

\cite{qiu_modeling_2015} & micro-level + macro-level & Debate forums & The probability of user 
choosing a stance  on an issue 
& users 4,994, issues 1,727arguments 39,549 (average 23 per issue)
\\ \hline

\cite{rui_weakly-guided_2017}     & (micro-level)& Online-fourms   & Stance(issue,posts)={pro,con}                                                                  & news articles from CNN and 4fourms 
                                                                  \\ \hline
\cite{darwish_predicting_2017}  & (micro-level)  & Twitter  & Stance(previous tweets,profiles,topic)={POS,NEG}  & Two data sets (not available)
Islands dataset(Arabic) [33,024 tweets| 2,464Users]
Islam dataset [7,203 tweets | 3,412 Users]		
 \\ \hline                                                    
\end{tabularx}}
\label{table:stance_prediction}
\end{table}

\begin{landscape}
\footnotesize\setlength{\tabcolsep}{2pt}
\begin{figure}
 \begin{adjustwidth}{-3cm}{-3cm}
\begin{longtable}
	 {|p{0.15\linewidth}|p{0.1\linewidth}| p{0.1\linewidth}| p{0.1\linewidth} |p{0.2\linewidth}|p{0.23\linewidth} |p{0.06\linewidth}|}	
        \hline
        Data-set  & Type & Application & Source of Data & Topics  & Annotation &Size \\ \hline
		
        Rumours dataset \cite{qazvinian2011rumor}& Claim-based & Rumours resolutions & Twitter & Obama, Airfrance, cellphone, Michelle, palin & Stance(users tweets,rumour's claim)=\{Confirm, Deny, Doubtful\} & 10,417 tweets                                        			                                                                                                                                                                                                                                 \\ \hline
		SemEval-Stance Detection Task A\cite{mohammad2016}  & Target
specific &General &Twitter & Atheism, Climate Change is a Real Concern, Feminist Movement, Hillary Clinton, and Legalization of Abortion & Stance(target, tweet)=\{Favor, Against, Neither\}            &  Training: 2,914  Testing: 1,249.  \\ \hline  
        
SemEval-Stance Detection Task B \cite{mohammad2016} & Target
specific &General & Twitter & Donald Trump  & Stance(target, tweet)=\{Favor, Against, Neither\}  &  707  labeled tweets 78,000 unlabeled \\ \hline
 
Emergent dataset \cite{ferreira_emergent:_2016} & Claim-based& Fact checking & variety of sources &  world and national U.S. news and technology stories.& Stance(claim Headline, article )=\{for, against, observing\}  &300 claims, 2,595 associated article headlines                                                                                                                                                                                   \\ \hline

SemEval Rumour stance classification -Subtask A - SDQC Support/ Rumour stance classification  \cite{derczynski_SemEval-2017_2017}& Claim-based & The Veracity of a rumour. & Twitter  & charliehebdo, ebola-essien, ferguson, germanwings-crash, ottawa shooting, prince-toronto, putinmissing, sydneysiege                                                                     &  Stance(originating rumourous, tweets reply)= \{deny, support, commenting\} &training: 4519 threads, Testing: 1049 threads\\ \hline

SemEval Rumour stance classification  Subtask B - Veracity prediction \cite{derczynski_SemEval-2017_2017}    & Claim-based & The Veracity of a
rumour. & Twitter & charliehebdo, ebola-essien, ferguson, germanwings-crash, ottawa shooting, prince-toronto, putinmissing, sydneysiege                                                                     &  Stance(originating rumourous, tweets reply)= \{deny, support, commenting\} &training: 297 threads , Testing: 28 threads\\ \hline

Fake News challenge 1 (FNC-1) {[}1{]}  & Target specific &Fake news detection & Derived from Emergent dataset & various news articles &  Stance(article headline, a body text)=\{ Agrees, Disagrees, Discusses, Unrelated\}   &49972 instances                                                                                                                  \\ \hline

Multi-Targets Stance Detection \cite{sobhani_dataset_2017}                                                                    & Multi-related-targets & General & Twitter   & Hillary Clinton-Sanders, Hillary Clinton-Trump and Cruz-Trump   &  Stance (multi-targets, tweet)=\{ Target A Favor|Against Target B\}&4,455 tweets     \\ \hline

Trump vs. Hillary dataset\cite{darwish2017trump} &  Multi-related-targets & General & Twitter & Hillary Clinton-Donald Trump & Stance (multi-targets, tweet)=\{ Target A Favor|Against Target B\} & 3,450 tweets\\ \hline

The Claim Polarity Dataset \cite{bar2017stance}  & Claim-based &Argument
construction & Claims are from Wikipedia  and motions from (IDEA)  website  & 5 topics selected randomly  & Stance(claim, motion)=\{Pro,Con\} & 2,394 claims.   \\ \hline
RumourEval 2019-Task A  & Claim-based &
The Veracity of a
rumour & Tweets  &  Various topics   & Stance(originating rumourous, tweets reply)= \{deny, support, commenting\} & 5216 training tweets  \\ \hline
 
Hyperpartisan News Detection SemEval2019  & Claim-based & Allegiance of news
  & News articles  &  various topics   & Stance(article, entity)=\{ prejudiced, non-hyperpartisan \}. & 750,000 articles  \\ \hline

X-stance  & Target
specific & Allegiance of news
  & comments from smartvote.ch  &  150 targets   & Stance(comment, target)=\{ support, against \}. & 67,000  comments  \\ \hline

\caption{Publicly available data-sets with stance annotations for stance classification}
  
\label{table:stance_Dataset}
\end{longtable}
\end{adjustwidth}
\end{figure}
\end{landscape}

\section{Discussion}

Social media provides a rich source for social studies to analyze public opinion, especially on controversial issue. While sentiment analysis has been used for decades to analyse public opinion on products and services, stance detection comes as the correspondent solution for analysing public opinion on political and social topics, where sentiment analysis fails to reflect support. Stance detection in social media presents an optimal technique for various applications and has been used heavily in interdisciplinary studies including politics, religion and humanities.     

It is worth noticing that small amount of work used stance to analyze the social issues in comparison with political topics. This is due to the controversial nature of the political topics which facilitates the data collection for the stance detection. 

The sentiment analysis has a minuscule effect on detecting the accurate stance on social media. This is due to the complexity of stance modeling on social media, which can be inferred based on various factors ranging from the textual cues to the social identity of the author. Nevertheless, there has been a noticeable misconception in the literature where the sentiment has been used solely to indicate the stance. These studies defined a linear relation between the against/in-favor stance and the negative/ positive sentiment. As the analysis in section \ref{sec:senti_analysis} shows that the relation between sentiment and stance is orthogonal.

Stance detection has been mostly approached using classification-based algorithm. This is mostly applied using supervised learning algorithms with huge dependency on human-annotated data. Consequently,  techniques such as transfer learning and unsupervised learning have been used to resolve the scarcity of the annotated data but with less attention from researchers compared to supervised methods. This scarcity reflected by the need to enrich the data with information related to the object of interest. For instance, to detect stances related to climate change, information related to global warming considered beneficial for stance detection in-order to cover the complete aspect of the topic. Other studies worked on overcoming this issue by using distant supervision to annotate the data without having to manually label the data. 

On the other end of the spectrum, few studies have dealt with stance prediction, which is mostly handled in relation with specific events. The main goal behind this kind of stance detection is to predicting the unexpressed views and to infer people standpoints toward an event in advance (pre-event). Therefore, stance prediction suits the analytical studies to examine the temporal effects of various events on public opinion such as candidates elections \cite{darwish2017trump}  or social phenomena as Islamophobic \cite{darwish_predicting_2017}. In this kind of studies, the dataset contains pre-event and post-event posts annotated with the user's stance before and after the event consequently. Thereby, predicting the stances is based on the user's past behavior which can be extracted from network features along with the post's content \cite{himelboim2013birds}. 

While there has been a large interest from the NLP community on developing effective approaches for stance detection~\cite{kuccuk2020stance} that mainly modeled the task as a text-entailment task, there is also large amount of work from the social computing and computational social science communities that showed the effectiveness of using user's interactions and network feature for stance detection and prediction. This shows that the task of stance detection is a multidisciplinary task that has been of interest to multiple computer science communities.

\subsection{Possible directions for moving forward}

The work on stance detection on social media has been growing over the last few years. More research gets published from multiple communities on the task and more applications get created. We believe that there are some gaps in the existing work and potential emerging directions that should receive more attention in the near future on this task. In the following, we discuss these directions.

    Some of the published work on transfer learning techniques for stance detection has shown some promising performance. The current techniques which used a topic based transfer learning achieved lower performance score in comparison with other techniques such as supervised learning techniques, as shown in section \ref{sec:best_algorithm}, however still promising. The transfer learning techniques for stance detection need a further enhancement by adapting a non-topical aspects to enhance the current state of this methodology on detecting the stance on social media, and potentially reaching the point to have comparable results with supervised on-topic stance detection. This might lead at some point to a general stance-classifier that can adapt to the topic of interest without the need to manually label new training data.  
    
    Several studies showed that stance detection models show consistent improvement on the overall performance when using the network features instead of just using content of post only. This kind of stance modeling draws more emphasis on the crucial role of users online behaviour and social attributes in the stance detection models. While network features are more effective, but they still require more computationally expensive, since they require gather additional information about users, which might be not highly practical in some cases. The need for further investigation in this direction is required to understand how to reach more effective, but at the same time highly efficient stance detection models that utilise network information. 

    Currently, the SemEval 2016 stance dataset \cite{mohammad2016} is the most popular benchmark dataset for target specific stance detection, and has served perfectly in many research publications, especially within the NLP community. Nevertheless, this dataset contains an average of only 500 records for each topic in the training set, which is quite limited. In addition, it is becoming old overtime, which means getting some of its tweets deleted, and thus failing to retrieve network information of their users for future research~\cite{aldayel_your_2019}.
    There is a real need for a new benchmark dataset for stance detection that provides a sufficient amount of data to train and test the stance detection models. In addition, most of the existing datasets are mainly focusing on social, political and religious issues. Having new datasets that covers additional domains would be of importance to explore the stance detection task for these new domains. Finally, creating benchmark datasets of multiple languages would be of interest, especially for the NLP community, where they focus on content-based stance detection approaches. 

    In general, as stance detection have been used heavily as social sensing technique on social media to study societal aspects ranging from politic, religion and social topics, this urges the need to incorporate more robust modeling of the stance on  social  by using network features for a more accurate analysis. As the use of textual entailment modeling shown to be sub-optimal when compared with network representation of the stance on social media section \ref{sec:analysis_NW_txt}. Despite the undeniable benefits of stance detection for various societal studies, privacy measures could be developed to better preserve the privacy of the social media users. These privacy methods can provide a green analysis methods for social sciences and humanities with minimal cost of redesigning the social media platforms. 



\section{Conclusion}
This survey covered the work on stance detection on social media and provided an overview of the current approaches to handle stance modeling. We initially presented the main difference between stance detection and sentiment analysis to remove any confusion between both techniques when used for measuring public opinion online. Later, we described the different types of targets when applying stance detection, namely: target-specific, multi-related-targets, and claim-based. Then we showed the existing work on detecting expressed stance or predicting unexpressed user's stance on future events. Later, the most used features for modeling stance and the different machine learning approaches used were discussed and compared, showing that network features are superior to content (textual features) for most of the studies on stance detection. In addition, supervised methods using SVM were found to be the most effective for different datasets. Also, the recent attempts for applying transfer learning and unsupervised learning for stance detection have promising results and is expected to be one of the main research direction in this area in the future.
We also discussed the different applications of stance detection, including fake news detection which is one of the most popular research topics in the recent couple of years. Finally, we summarise the existing resources on stance detection, including datasets and most effective approaches.

Our final discussion on the survey listed the potential future directions in this area, including the need for having new datasets and the importance of using NLP and social computing approaches for better stance detection on social media.

\bibliographystyle{elsarticle-num}

\bibliography{surveyref1}
\newpage
\appendix 
\section{Appendix A: Stance Detection Work} 
This appendix provides list of the most recent stance prediction and detection work on social media

\small
\begin{longtable} 
{|p{0.12\linewidth}|p{0.3\linewidth}| p{0.25\linewidth} |p{0.25\linewidth} |c|}

  \hline
   Study & Features & ML &  Dataset   \\
  \hline\hline
 \cite{darwish_predicting_2017} &  Content Features
(Hashtags, Text); Profile Features (Description, Name, Location); Network Features (Mention,
Reply, Retweet) & SVM & Islamophobic dataset (Twitter) [Not available]   \\ \hline
 \cite{magdy__2016} &  Content Features
(Hashtags, Text); Profile Features (Description, Name, Location); Network Features (Mention,
Reply, Retweet) & SVM & Islamophobic dataset (Twitter) [Not available] \\ \hline
 \cite{darwish_improved_2017} &  Content Features(Text); Interaction Elements(retweeted
accounts,  used hashtags, mentioned
accounts (MEN), shared URLs (URL)); User Similarity  & SVM & Islands Dataset and Islamophobic dataset (Twitter)   [Not available] \\ \hline
\cite{lahoti2018joint} &   A combination of
network and content  & Non-negative matrix factorization  &  dataset covered Three  controversial topics:gun control,abortion and obamacare (Twitter)  [Not available] \\ \hline
\cite{gottipati2013predicting}&  similarity between users  &  Probabilistic Matrix Factorization  &  1000 user profile of Democrats and Republicans (debate.org)  [Not available] \\ \hline
 
\cite{rui_weakly-guided_2017} &  post level interaction and user level interaction  &  Stance-based Text Generative Model with Rule-based User-User
Interaction Links & CNN dataset, 4Forums and IAC discussion forum  [Not available] \\ \hline
\caption{Work in stance prediction} 
\label{table:allinfo}
\small
\end{longtable}


\begin{longtable} 
{|p{0.12\linewidth}|p{0.08\linewidth}| p{0.25\linewidth}| p{0.25\linewidth} | p{0.2\linewidth} |c|}

  \hline
   Study & Task & Features & ML &  Dataset  \\
  \hline\hline
 \cite{aldayel_your_2019} & target-specific &    NW features & SVM& SemEval-2016 shared task 6 [Available]\\\hline
 
  \cite{lynn2019tweet}  & target-specific &   NW (followee) & RNN & SemEval-2016 shared task 6 [Available]\\\hline

 \cite{siddiqua_tweet_2019-1}  & target-specific &  Content & Nested  LSTMs &  SemEval-2016 shared task 6 [Available]\\\hline

 \cite{sun2018stance}  & target-specific &  Content & Hierarchical Attention NN & SemEval-2016 shared task 6 [Available]\\\hline

  \cite{siddiqua2018stance}  & target-specific &  Content &  SVM Tree Kernel & SemEval-2016 shared task 6 [Available]\\\hline
 
\cite{wei2019topic} & target-specific &  Content+Sentiment lexicon & BiLSTM  &  SemEval-2016 shared task 6 [Available]\\\hline

 \cite{wei2019topic}  & target-specific  &  Content+Noisy stance labeling + Topic Modeling & BiGRU & SemEval-2016 shared task 6 [Available]\\\hline

 \cite{ebner2019bag}  & target-specific &   words embedding & Deep averaging network &SemEval-2016 shared task 6 [Available]\\\hline
 
 \cite{liu_iucl_2016} &  target-specific &   bag-of-words and word vectors (GloVe  and word2vec) & gradient boosting decision trees and SVM and
merge all classifiers into an ensemble method &   SemEval-2016
shared task 6  [Available] \\ \hline
 
 \cite{dias_inf-ufrgs-opinion-mining_2016} & target-specific &   n-gram and sentiment  &  SVM & SemEval-2016
shared task 6  [Available] \\ \hline

\cite{dias_inf-ufrgs-opinion-mining_2016} & target-specific &   n-gram and sentiment  &  SVM & SemEval-2016
shared task 6  [Available] \\ \hline

\cite{igarashi_tohoku_2016} & target-specific &   Reply, BagOfWord, BagOfDependencies, POS tags Sentiment WordNet, Sentiment Word Subject, Target Sentiment and Point-wise Mutual Information &  CNN  & SemEval-2016
shared task 6  [Available] \\ \hline

\cite{igarashi_tohoku_2016} & target-specific &   Reply, BagOfWord, BagOfDependencies, POS tags Sentiment WordNet, Sentiment Word Subject, Target Sentiment and Point-wise Mutual Information &  CNN  & SemEval-2016
shared task 6  [Available] \\ \hline
\cite{augenstein_stance_2016} & target-specific & word2vec  &  Bidirectional LSTMs   & SemEval-2016
shared task 6   [Available] \\ \hline
\cite{krejzl_uwb_2016} & target-specific & hashtags, n-grams, tweet length, Part-of-speech, General Inquirer, entity-centered
sentiment dictionaries, Domain Stance Dictionary  & Maximum entropy classifier & 
SemEval-2016 shared task 6  [Available] \\ \hline
\cite{ebrahimi_joint_2016} & target-specific & n-gram and sentiments & discriminative and generative models & SemEval-2016 shared task 6   [Available] \\ \hline
\cite{wei_pkudblab_2016} & target-specific & Google news word2vec and hashtags& CNN  &  SemEval-2016 shared task 6   [Available] \\ \hline
\cite{zarrella_mitre_2016} & target-specific & word2vec hash-tags  & LSTM  &  SemEval-2016 shared task 6 [Available] \\ \hline
\cite{rajadesingan2014identifying} & target-specific & unigrams, bigrams and
trigrams & Naive Bayes & hotly contested gun reforms debate from April 15th, 2013 to April 18th, 2013.  [Available]  \\ \hline
\cite{zhou_connecting_2017} & target-specific &  word embeddings  & Bi-directional GRU-CNN & SemEval-2016 shared task 6   [Available] \\ \hline
\cite{vijayaraghavan_deepstance_2016} & target-specific &    word embeddings & convolutional neural networks(CNN) & SemEval-2016 shared task 6   [Available]  \\ \hline

\cite{elfardy2016cu} & target-specific &   Lexical Features,  Latent Semantics, Sentiment,  Linguistic Inquiry , Word Count and Frame Semantics features &  SVM & SemEval-2016 shared task 6 [Available] \\ \hline

 \cite{lai2016friends} & target-specific &  sentiment, Opinion
target, Structural Features(Hashtags, Mentions, Punctuation marks), text-Based Features(targetByName, targetByPronoun, targetParty, targetPartyColleagues) &  Gaussian Naive Bayes classifier  & Hillary Clinton and Donald Trump dataset [Not available] \\ \hline

 \cite{sobhani_dataset_2017} & multi-target &  word vectors  &  bidirectional
recurrent neural network (RNN) & Multi-Target Stance dataset [Available] \\ \hline

\cite{siddiqua2019tweet} & multi-target & content of tweets & Multi-Kernel Convolution and Attentive LSTM & Multi-Target Stance dataset [Available] \\ \hline

\cite{bar2017stance} &  claim-based & Contrast scores   &  random forest and SVM  &  The claim polarity dataset claims are from Wikipedia and motions f are from rom (IDEA)(on-line forums)   [Available] \\ \hline
\cite{aker_simple_2017} &  claim-based  &  Linguistic, message-based, and topic-based such as (Bag of words, POS tag, Sentiment, Named entity and others & Random Forest, Decision tree and Instance Based classifier (K-NN) &  RumourEval and  PHEME datasets   [Available]\\ \hline

\cite{hamidian2015rumor} & claim-based &    tweet content, Unigram-Bigram Bag of Words, Part of Speech, Sentiment, Emoticon, Named-Entity Recognition , event, time, Reply, Re-tweet, User ID, Hashtag, URL  & decision trees & \cite{qazvinian2011rumor} [Available] \\ \hline
\cite{aker_simple_2017} & claim-based &  BOW,Brown Cluster, POS tag, Sentiment, Named entity, Reply, Emoticon, URL, Mood, Originality score, is User Verified(0-1),Number Of Followers, Role  score, Engagement score, Favourites score and other tweets related  features & decision tree, Random Forests and Instance Based classifier    & RumourEval dataset \cite{derczynski_SemEval-2017_2017} and
the PHEME dataset \cite{derczynski2015pheme} [Available] \\ \hline
\cite{zubiaga_discourse-aware_2018} & claim-based &   Word2Vec, POS, Use of negation, Use of swear words, Tweet length, Word count, Use of question mark, Use of exclamation mark,Attachment of URL and other contextualized features  & Linear CRF and tree CRF , a Long Short-Term Memory (LSTM)  & PHEME dataset \cite{derczynski2015pheme} and Rmour dataset associated with eight events corresponding to breaking news events \cite{zubiaga_analysing_2016} [Available] \\ \hline
\cite{kochkina_turing_2017} & claim-based &   word2vec, Tweet lexicon (count of negation words and  count of swear words),Punctuation, Attachments,Relation to other tweets, Content length and Tweet role (source tweet of a conversation) & branch-LSTM , a neural network architecture that uses layers  of  LSTM units  & Rumoureval dataset \cite{derczynski_SemEval-2017_2017}  [Available] \\ \hline

\caption{Work in stance classification} 
\label{table:allinfo2}
\small
\end{longtable}

\end{document}